\documentclass[10pt, conference, letterpaper]{IEEEtran}

\usepackage{graphics,graphicx,amsmath}
\usepackage{color}
\usepackage{epsfig}
\usepackage{subfig}

\newfont{\mbb}{msbm10 scaled 1100}

\definecolor{red}{rgb}{0.9,0.0,0.1}

\definecolor{blue}{rgb}{.15,.4,.8}

\definecolor{green}{rgb}{.25,.8,.25}

\definecolor{orange}{RGB}{255,128,0}

\definecolor{dgreen}{RGB}{0,151,0}

\def\b0{{\bf 0}}

\title{Systematic Network Coding with the Aid of a Full-Duplex Relay}
\author{\authorblockN{Giuliano Giacaglia\authorrefmark{1}, Xiaomeng Shi\authorrefmark{1}, MinJi Kim\authorrefmark{1}, Daniel E. Lucani\authorrefmark{2}, and Muriel M\'edard\authorrefmark{1}\vspace*{.2cm}}
\authorblockA{\authorrefmark{1}Research Laboratory of Electronics, Massachusetts Institute of Technology, Cambridge, MA 02139} 
\authorblockA{\authorrefmark{2}Instituto de Telecomunica\c c\~oes, DEEC Faculdade de Engenharia, Universidade do Porto, Portugal\\ Email: \{giu, xshi, minjikim, medard\}@mit.edu, dlucani@fe.up.pt}\vspace*{-.7cm}}

\begin{document}

\maketitle

\begin{abstract}
A characterization of systematic network coding over multi-hop wireless networks is key towards understanding the trade-off between complexity and delay performance of networks that preserve the systematic structure. This paper studies the case of a relay channel, where the source's objective is to deliver a given number of data packets to a receiver with the aid of a relay. The source broadcasts to both the receiver and the relay using one frequency, while the relay uses another frequency for transmissions to the receiver, allowing for a full-duplex operation of the relay.
We analyze the decoding complexity and delay performance of two types of relays: one that preserves the systematic structure of the code from the source; another that does not.
A systematic relay forwards uncoded packets upon reception, but transmits coded packets to the receiver after receiving the first coded packet from the source.
On the other hand, a non-systematic relay always transmits linear combinations of previously received packets.
We compare the performance of these two alternatives by analytically characterizing the expected transmission completion time as well as the number of uncoded packets forwarded by the relay. Our numerical results show that, for a poor channel between the source and the receiver, preserving the systematic structure at the relay (i) allows a significant increase in the number of uncoded packets received by the receiver, thus reducing the decoding complexity, and (ii) preserves close to optimal delay performance.

\end{abstract}

\section{Introduction}\label{sec:intro}
Introduced in~\cite{ahlswede}, network coding constitutes an attractive technique to enable cooperative communication in multi-hop networks. It allows intermediate nodes to send a function of previously received packets to downstream nodes instead of restricting their operation to a store-and-forward strategy. Linear codes over a network are known to be sufficient to achieve multicast capacity~\cite{Li03}, while generating these linear codes at random achieves capacity with high probability~\cite{Ho06}. An important argument against this last approach, called random linear network coding (RLNC), is that its decoding complexity scales as $O(n^3)$ for recovering $n$ data packets. A simple approach to maintain delay performance while reducing decoding complexity is to code systematically. In systematic network coding, the source first sends the original data packets, uncoded; it then transmits random linear combinations of those original packets using RLNC. A large body of literature has considered systematic network coding in different applications focusing mostly on single-hop networks. To the best of our knowledge, only the work in~\cite{Shrader09} considered the case of multi-hop networks, providing an analysis of the packet loss rate.

\begin{figure}[t!]
\centering
\includegraphics[width = 0.25\textwidth]{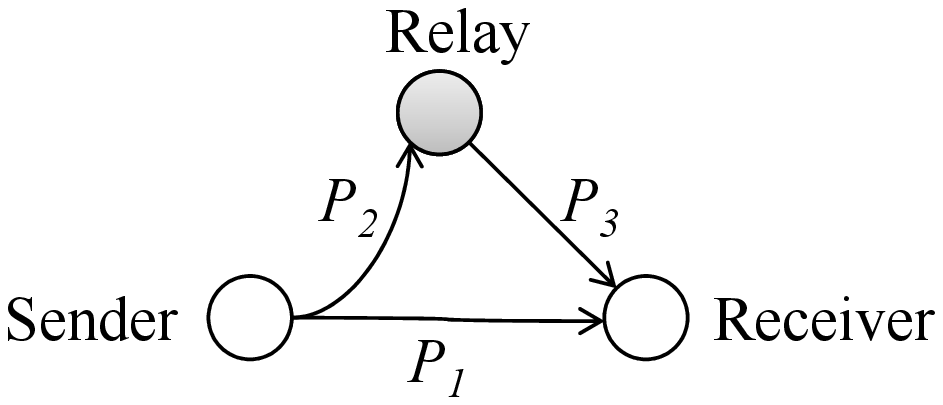}\vspace*{-.2cm}
\caption{The topology of the relay network with packet erasure probabilities $P_1$, $P_2$ and $P_3$. The source broadcasts to both relay and receiver; however, the relay and receiver experiences losses independently.}\label{fig:topology}\vspace*{-.4cm}
\end{figure}

Ref.~\cite{Yazdi09} considered systematic network coding as a MAC layer mechanism for WiMAX, showing that it achieves optimum performance for delay sensitive applications with overheads similar to a purely RLNC mechanism proposed in~\cite{Jin08}.
Heide et al.~\cite{Heide09} characterized the delay performance as well as the capabilities of mobile devices implementing  systematic network coding with GF(2). The work in~\cite{Lucani10} studied the expected computational complexity of systematic network coding, considering the loss probability of the channel. Ref.~\cite{Lucani10} also
showed that systematic network coding, even with small field sizes, can have close to or the same performance  in terms of the mean completion time as RLNC with a large field size. This means that a systematic approach enables the system to rely on simpler and fewer operations to achieve close to optimal delay performance.
Online mechanisms that leverage systematic network coding ideas have been considered in~\cite{Barros09}, focusing on improvements in decoding delay.

This paper proposes delay optimal and sub-optimal mechanisms that attempt to preserve the systematic structure introduced at the source in multi-hop wireless networks. In particular, our work considers a simple, yet relevant topology consisting of one sender, one relay, and one receiver. The relay is assumed to have an infinite queue for received packets, which allows the generation of random linear coded packets at each time slot if needed. We provide two relay-centered schemes, called \textit{systematic} and \textit{non-systematic}, and characterize the completion time as well as decoding complexity performance of these schemes using a Markov chain model inspired by the work in~\cite{Nistor11}. Our work is different from~\cite{Shrader09}, since we focus on rateless schemes and characterize the delay-complexity trade-off, instead of focusing on determining the packet loss rate of a fixed-rate scheme.

This paper is organized as follows. In Section~\ref{sec:model}, we outline the system model. In Sections~\ref{sec:non-sys} and~\ref{sec:sys}, we analyze the mean completion time in terms of the transition probabilities of a Markov Chain for the \emph{non-systematic} and \emph{systematic} relays, respectively. In Section~\ref{sec:decoding}, we discuss the effect on decoding complexity. In Section \ref{sec:simulations}, we present numerical results to support our analysis and conclude in Section~\ref{sec:conclusions}. 
\section{System Model}\label{sec:model}
A single source has $M$ data packets, denoted $\mathbf{p}_1, \mathbf{p}_2, ... \mathbf{p}_M$, to send to a receiver. The sender broadcasts to both the receiver and the relay. The relay is assumed to have an infinite buffer, keeping all received packets in its memory until the connection terminates. The relay is assumed to operate in full-duplex mode, transmitting to the receiver on a frequency different from the source, thus capable of listening to the source while transmitting to the receiver. At each time step, the relay sends a packet to the receiver as long as its queue is not empty.

Figure~\ref{fig:topology} illustrates the network topology. The arrows indicate possible paths that a packet can traverse. The broadcast link is represented by an hyperedge. For each wireless link, we assume packet erasure channels with static loss probabilities, i.e. a packet is either received or lost completely with a given probability. We denote the packet erasure rates for channels between source-receiver, source-relay, and relay-receiver by $P_1$, $P_2$, and $P_3$, respectively. Time is assumed to be slotted, with each packet transmission requiring one time unit. Observe that the relay can only transmit a new packet to the receiver after it receives the packet successfully from the source. Hence from the relay's perspective, it is always operating one time unit behind the source. In this paper, we ignore the effect of this additional unit time lag, since its effect on overall system performance is negligible if $M$ is large.

At the sender, systematic network coding is performed through two stages of transmissions. In the first stage, the sender broadcasts the $M$ original data packets in order. In the second stage, the source generates coded packets for transmission. 
For each packet, the source generates a linear combination $\sum_{i=1}^{M}c_i\mathbf{p}_i$ where $c_i$'s are coding coefficients. The coefficients are concatenated into a vector $(c_1,c_2,\ldots,c_M)$, then attached to the linear combination for transmission. During the first stage of transmission, uncoded packets $\mathbf{p}_i$ are sent with a unit coefficient vector $(0,0,\ldots,1,\ldots,0)$, where the index of $1$ is $i$. 
During the second stage, the source generates coding coefficients $c_i$'s randomly.

The transmission process terminates when the receiver acknowledges the receptions of $M$ degrees of freedom (dof), where dofs represent linearly independent packets received. The acknowledgment process is assumed to incur no delay. In our analysis in Sections~\ref{sec:non-sys} and \ref{sec:sys}, we assume that RLNC is performed in a field of size $q$, where $q$ is sufficiently large, such that the effect of field size on generating linearly dependent packets is negligible. Under this assumption, each packet from the source to the receiver is innovative, in the sense that if received successfully, it becomes a new dof at the receiver. In our discussion in Section~\ref{sec:simulations}, we present numerical results for $q=\infty$ and simulation results for $q=2^4$. From a throughput point of view, systematic coding at the source is equivalent to non-systematic coding. However, the use of systematic codes can lead to significant reductions in encoding and decoding complexities.

We assume in this paper that the source always transmits systematically, while the relay may operate systematically or non-systematically.
\begin{itemize}
\item \emph{Non-systematic}: regardless of the stage, the relay always sends a linear combination of packets in its queue.
\item \emph{Systematic}: during the first stage, the relay forwards uncoded packets  received from the sender without re-encoding. The relay transmits each uncoded packet only once, and stores it in memory. During the second stage, the relay transmits linear combinations of all packets in its memory, regardless of whether they are coded or uncoded.
\end{itemize}
If the relay has not received any packets successfully and its memory is empty, a null packet is assumed to be transmitted by the relay. 
\section{Non-systematic Relay}\label{sec:non-sys}
To analyze the time required to send $M$ dofs from the source to the receiver with a non-systematic relay, a Markov chain model can be established. This approach is similar to that of \cite{lucani_fieldsize}, which characterizes transmission delays when RLNC is applied over a single link.

We define the state of the network by a 3-tuple $(i,j,k)$, where $i$ and $j$ represent the number of dofs at the receiver and the relay, respectively, while $k$ represents the number of dofs shared by these two nodes. Since there are in total $M$ dofs, $(i,j,k)$ is valid if and only if $i+j-k \leq M$, $i\leq M$, $j\leq M$, and $ 0 \leq k \leq \min(i,j)$. Transmission starts in $(0,0,0)$, and terminates in $(M,j,k)$. There is an partial order of states $(i,j,k) \preceq (i',j',k')$ according to the validity condition of state transitions: $i\leq i'$, $j\leq j'$ and $k\leq k'$.

\begin{figure}[t!]
\centering
\includegraphics[width=0.22\textwidth]{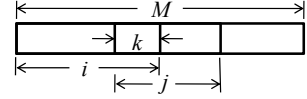}
\caption{A representation of the state $(i,j,k)$. Given $M$ degrees of freedom (dof) at the source, the receiver and the relay have $i$ and $j$ dofs, respectively; they also share $k$ dofs.}\vspace*{-.3cm}
\label{fig:state}
\end{figure}

Let $P_{(i,j,k) \rightarrow (i',j',k')}$ be the probability of transiting from state $(i,j,k)$ to state $(i',j',k')$ when one packet is transmitted by the source. Let $T_{(i,j,k)}$ be the expected amount of time to reach a terminating state from $(i,j,k)$. Under the slotted transmission model, 
$T_{(i,j,k)}$ can be defined recursively as
\begin{align}
T_{(i,j,k)} & = 1 +\sum_{(i,j,k) \preceq (i',j',k')}P_{(i,j,k) \rightarrow (i',j',k')} T_{(i',j',k')}\,. \label{eq:Tijk}
\end{align}
For terminating states, $T_{(M,j,k)}=0$.

\subsection{Transition Probabilities}

\begin{figure}[tbp]
\begin{center}
\includegraphics[width=0.49\textwidth]{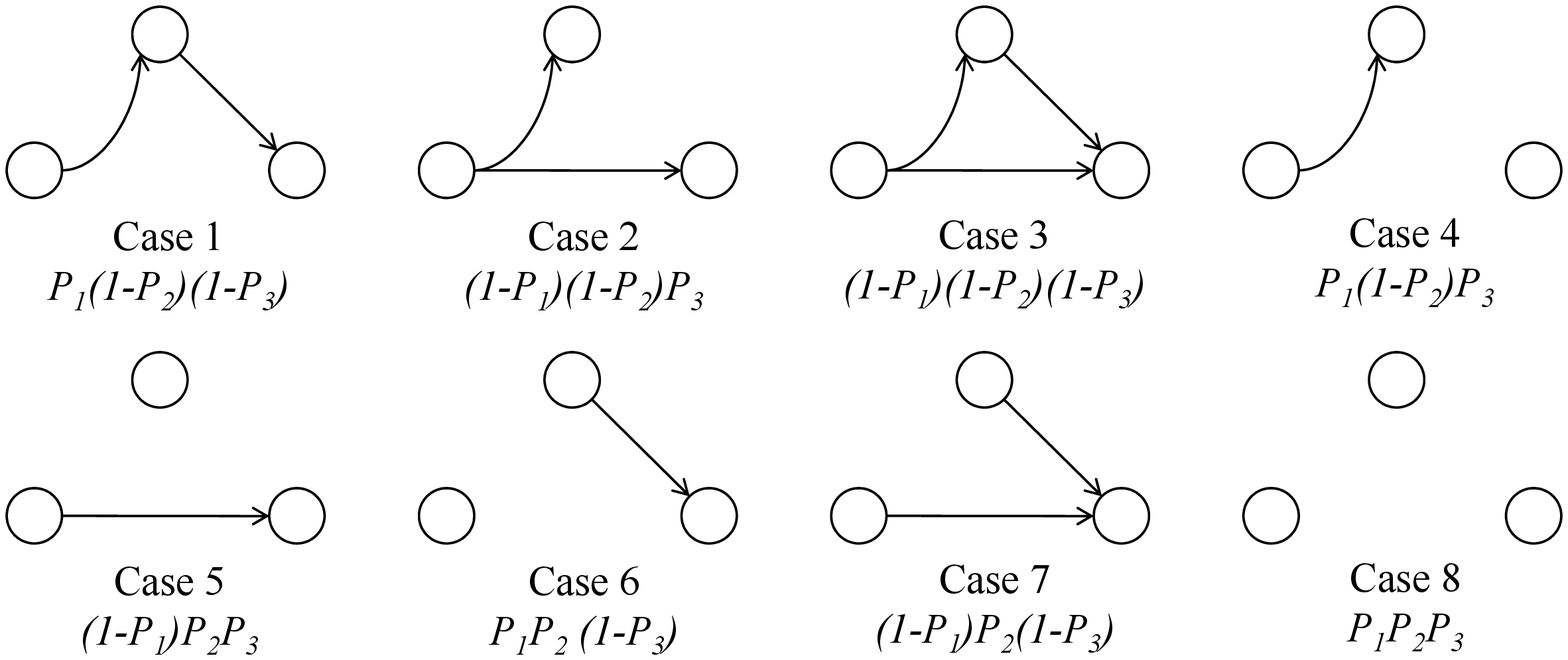}
\caption{Possible transmission patterns and corresponding probabilities.}
\label{fig:statetransitions}\vspace*{-.4cm}
\end{center}
\end{figure}

For a non-systematic relay, Eq.~(\ref{eq:Tijk}) shows that to find the completion time, we only need to compute state transition probabilities $P_{(i,j,k) \rightarrow (i',j',k')}$, assuming that a single packet, coded or uncoded, is transmitted from the source. Depending on the values of $i$, $j$, and $k$, we can divide the computation of $P_{(i,j,k) \rightarrow (i',j',k')}$ into 5 different cases as listed below. For each case, all possible state transitions are considered, as illustrated by Figure~\ref{fig:statetransitions}. In Figure \ref{fig:statetransitions}, an arrow represents a successful transmission of one packet from a node to another. 

We denote $(\delta i,\delta j,\delta k)$ to be the updates to $i$, $j$, and $k$, where
\begin{align*}
i' &=\min(i+\delta i, M)\,, \quad j' = \min(j + \delta j, M)\,, \quad \text{ and}\\
k' &= \min(k+\delta k,M,i+ \delta j, k + \delta k)\,.
 \end{align*}
 The transition probability $P_{(i,j,k) \rightarrow (i',j',k')}$ can be found by adding the probabilities of cases given in Figure~\ref{fig:statetransitions} corresponding to a given value of $(\delta i,\delta j,\delta k)$.

\subsubsection{$i+j-k=M$ and $k<\min(i,j)$}\label{sec:case_a} since $i+j-k=M$, the combined knowledge at the relay and the receiver is equal to $M$. Therefore, $\delta k = \delta i + \delta j$, implying that any additional dof received at the receiver/relay contributes to the common knowledge $k$ between them.
A dof new to the relay is already in the the knowledge space of the receiver, and vice versa. Thus, if both the relay and the receiver successfully receive a packet from the source, $k$ increments by two. 

Since $k < \min(i, j)$, the relay has at least one dof not known to the receiver; otherwise, $k$ cannot be strictly less than both $i$ and $j$. In other words, a packet from the relay to the receiver is innovative to the receiver.


The probability $P_{(i,j,k) \rightarrow (i',j' j,k')}$ is given by:
	    \begin{itemize}
            \item $(\delta i,\delta j,\delta k)=(1,1,2)$: Cases 1, 2.

                In Case 1, $\delta j = 1$ since the relay receives a  packet from the source; $\delta i = 1$ since the receiver receives a coded packet from the relay.                 In Case 2, the relay and the receiver both receive the same coded packet from the source. This packet is innovative to both nodes.

            \item $(\delta i,\delta j,\delta k)=(2,1,3)$: Case 3. 
            
               $\delta j = 1$ since the relay receives a coded packet from the source. $\delta i = 2$ since the receiver receives two packets, one from the source and one from the relay. Thus, $\delta k$ can increment by 3, since the single broadcast from the source increases $k$ by 2. 

            \item $(\delta i,\delta j,\delta k)=(0,1,1)$: Case 4.

            \item $(\delta i,\delta j,\delta k)=(1,0,1)$: Cases 5, 6.

            \item $(\delta i,\delta j,\delta k)=(2,0,2)$: Case 7.

	    \item $(\delta i,\delta j,\delta k)=(0,0,0)$: Case 8.
        \end{itemize}

\subsubsection{$i+j-k=M$ and $k=i<M$, then $j=M$} the relay already has all dofs. Hence, even if the relay receives a packet from the source, $j$ does not change. The relay can be considered as an additional source. As a result, any packet from the source or the relay is innovative to the receiver. Therefore, $\delta j = 0$ and $\delta i = \delta k$ for all cases.

The probability $P_{(i,j,k) \rightarrow (i' i,j',k')}$ is given by:
	    \begin{itemize}
            \item $(\delta i,\delta j,\delta k)=(1,0,1)$: Cases 1, 2, 5, 6.
            \item $(\delta i,\delta j,\delta k)=(2,0,2)$: Cases 3, 7.
            \item $(\delta i,\delta j,\delta k)=(0,0,0)$: Cases 4, 8.
	    \end{itemize}

\subsubsection{$i+j-k=M$ and $k=j$, then $i=M$} the receiver has received all dofs. Therefore, transmission is completed and the state is absorbing, i.e. $P_{(i,j,k) \rightarrow (i, j, k)} = 1$.

\subsubsection{$i+j-k <M$, and $k < \min(i,j)$ or $k = i < j$} since $k<j$, the relay has at least one dof which is not shared with the receiver. Therefore, a coded packet from the relay is innovative to the receiver. The key difference between this scenario and that of Section \ref{sec:case_a} is that the equality $\delta i + \delta j = \delta k$ does not necessarily hold here. The condition $i + j - k< M$ implies that the combined knowledge at the receiver and the relay is less than $M$. A coded packet from the source is innovative to both the relay and the receiver. If received successfully by both nodes, $k$ increments by one.

The probability $P_{(i,j,k) \rightarrow (i',j',k')}$ is given by:
    \begin{itemize}
        \item $(\delta i,\delta j,\delta k)=(1,1,1)$: Cases 1, 2.

            In Case 1, $\delta k = 1$ since the coded packet from the relay to the receiver is outside of the receiver's knowledge space. In Case 2, $\delta i = \delta j =\delta k = 1$ since the source transmits the same dof to both the relay and the receiver. 

        \item $(\delta i,\delta j,\delta k)=(2,1,2)$: Case 3.
	
	    Unlike in Section~\ref{sec:case_a}, $\delta k=2$ here since $i+j-k <M$ and the same new dof is transmitted from the source to the relay and the receiver. 

        \item $(\delta i,\delta j,\delta k)=(0,1,0)$: Case 4.

            $\delta k = 0$ since the packet from the source to the relay is not in the knowledge space of receiver.

        \item $(\delta i,\delta j,\delta k)=(1,0,0)$: Case 5.

            $\delta k = 0$ since the packet from the source to the receiver is not in the knowledge space of the relay.

        \item $(\delta i,\delta j,\delta k)=(1,0,1)$: Case 6.
        \item $(\delta i,\delta j,\delta k)=(2,0,1)$: Case 7.

            Since the packet from the source to the receiver is not within the knowledge space of the relay, this packet does not count towards $k$, resulting in $\delta k = 1$. 
            
        \item $(\delta i,\delta j,\delta k)=(0,0,0)$: Case 8.
    \end{itemize}

\subsubsection{$i+j-k <M$ and $k=j\leq i$}
since $k = j$, all dofs at relay are already known at the receiver. For $i$ to increase, a new dof needs to be delivered from the source directly or indirectly to the receiver. Therefore, receiving a packet at the receiver from the relay does not increase $i$ or $k$ unless the source has successfully transmitted a new packet to the relay. Furthermore, if the receiver receives packets from both the source and the relay,  then $\delta i = 1$.

Therefore, $P_{(i,j,k) \rightarrow (i',j',k')}$ is given by:

	    \begin{itemize}
            \item $(\delta i,\delta j,\delta k)=(1,1,1)$: Cases 1, 2, 3.
            \item $(\delta i,\delta j,\delta k)=(0,1,0)$: Case 4.
            \item $(\delta i,\delta j,\delta k)=(1,0,0)$: Cases 5, 7.
            \item $(\delta i,\delta j,\delta k)=(0,0,0)$: Cases 6, 8.
        \end{itemize}

%

\subsection{Mean Transmission Completion Time}

In the above analysis, we assume that the relay always codes packets in its queues. Since the system initiates in state $(0,0,0)$, the mean completion time for a non-systematic relay network is given by
\begin{align}
T_{non-sys}=T_{(0,0,0)},\label{eq:TNS}
\end{align}
where $T_{(0,0,0)}$ is defined recursively using Equation \eqref{eq:Tijk}. 
\section{Systematic Relay}\label{sec:sys}
In Section \ref{sec:non-sys} (non-systematic relay), we do not consider the two stages separately because the relay operates in the same non-systematic fashion in both stages. However, in the systematic relay case, the relay behaves differently in the two stages. In the first stage, whenever the relay receives an uncoded packet $\mathbf{p}_i$, it forwards $\mathbf{p}_i$ to the receiver and keeps $\mathbf{p}_i$ in its memory. Since no feedback is generated until the receiver is able to decode all $M$ packets, the relay does not attempt to transmit any uncoded packet more than once. This is done to reduce unnecessary repetitions to the receiver.

Therefore, if a transmission attempt of an uncoded packet from the relay to the receiver fails, this particular dof will not be re-transmitted by the relay until the second stage. In other words, the relay functions in a ``memoryless'' fashion during the first stage. By comparison, a non-systematic relay always combines contents of its memory, enabling the re-transmission of a particular dof even during the first stage. This setup enables a smaller mean transmission completion time for the non-systematic relay, at the cost of decoding complexity.

Similar to Section \ref{sec:non-sys}, the state of the network is defined by the 3-tuple $(i,j,k)$, where $i$ and $j$ represent the numbers of dofs at the receiver and the relay, respectively, while $k$ represents the number of dofs shared by these two nodes. Transmission initiates in $(0,0,0)$ and terminates in $(M,j,k)$.

\subsection{Transition Probabilities}
Let $Q_{(0,0,0) \rightarrow (i,j,k)}$ represent the probability of being in state $(i,j,k)$ at the end of the first stage of transmission. Whether an uncoded packet is delivered successfully from the source to the receiver during the first stage can be viewed as the result of three independent Bernoulli trials, with failure probabilities $P_1$, $P_2$ and $P_3$. 
Thus, when $M$ uncoded packets are sent by the source, $Q_{(0,0,0) \rightarrow (i,j,k)} = Q_a Q_b Q_c(P_1P_2)^{M-i-j+k}$, where
\begin{align*}
Q_a &=  {M \choose i-k} ((1-P_1)P_2)^{i-k}, \\
Q_b & = {M-i+k \choose j-k} ((1-P_2)P_1P_3)^{j-k}, \quad \text{and} \\
Q_c & = {M-i-j+2k \choose k}((1-P_2)(1-P_1P_3))^k.
\end{align*}

\noindent $Q_a$ represents the probability that from the $M$ uncoded packets sent by the source, $i-k$ are received by the receiver only; $Q_b$ represents the probability that from the remaining $M-(i-k)$ packets, $j-k$ are received by the relay only; $Q_c$ represents the probability that from the remaining $M-i-j+2k$ packets, $k$ are received by both the relay and the receiver; the factor $(P_1P_2)^{M-i-j+k}$ represents the probability that $M-i-j+k$ of the transmitted packets are lost completely, received by neither the relay nor the sink.

Once source starts the second stage, the system behaves exactly like that of the non-systematic relay in Section \ref{sec:non-sys}. As a result, we have the same state transition probabilities $P_{(i,j,k) \rightarrow (i',j',k')}$ during the second stage.

\subsection{Mean Transmission Completion Time}
Taking into account of the first stage, where the source broadcasts uncoded packets and the relay operates systematically, the expected completion time can be computed as
\begin{align}
T_{sys} & = M+\sum_{(0, 0, 0) \preceq (i,j,k)}Q_{(0,0,0) \rightarrow (i,j,k)}T_{(i,j,k)}\,,\label{eq:TS}
\end{align}
\noindent where $T_{(i,j,k)}$ is given by Equation \eqref{eq:Tijk}.

\section{Effect of Systematic Relay on Decoding}\label{sec:decoding}
To compare the complexity of the systematic and non-systematic relaying schemes, observe that decoding complexity at the receiver is directly proportional to $M-U$, where $U$ is the number of uncoded packets received by the receiver. If $U=M$, no decoding operation is required. If $U=0$, on the other hand, all $M$ dofs received at the sink are coded. In order to decode the original packets, the receiver first needs to back substitute the uncoded information to the coded information, which requires $O(U(M-U))$ operations; next, the receiver has to perform Gaussian elimination, which requires $O((M-U)^3)$ operations. Therefore, the overall decoding complexity is on the order of $O((M-U)^3) + O( U(M-U))$.

Since the source always operates in a systematic manner, the benefit of using a systematic relay is quantified by the number of uncoded packets successfully delivered through the relay to the receiver while not being delivered from the source to the relay directly. For example, if the source could already deliver an uncoded packet to the receiver directly, the relay having also delivered the same uncoded packet to the receiver cannot be considered a gain. However, if it is the case that the source fails to deliver an uncoded packet directly while the relay is able to do so successfully, this should be considered a gain in using a systematic relay.

We denote the number of uncoded dofs received at the receiver in the systematic and non-systematic relay cases as $U_{sys}$ and $U_{non-sys}$, respectively. The effect of using a systematic relay on decoding complexity is measured by the difference, $U_{sys} - U_{non-sys}$. For example, let $M=6$, and assume that the receiver is ready to decode. 
In order to decode, the receiver needs to use Gaussian elimination to invert $C$, where $C$ is the coding matrix generated by concatenating the coding coefficient vectors $(c_{j, 1}, c_{j, 2}, ..., c_{j, M})$ attached to the $j$-th received packet. For instance, $C$ could be of the following form when the relay operates systematically:
\begin{align*}
\scriptsize
C = \left(
\begin{array}{c c c c c c}
 1& 0 & 0 & 0 & 0 & 0 \\
\hline
 0& 0 & 1 & 0 & 0 & 0 \\
 0& 0 & 0 & 1 & 0 & 0 \\
 0& 0 & 0 & 0 & 0 & 1 \\
\hline
 c_{5,1} & c_{5,2} & c_{5,3} & c_{5,4} & c_{5,5} & c_{5,6} \\
 c_{6,1} & c_{6,2} & c_{6,3} & c_{6,4} & c_{6,5} & c_{6,6} \\
\end{array}
\right)\,,
\end{align*}
where $c_{j,k}$ are random coefficients. In this example, four of the six received packets are uncoded. Suppose that the first uncoded packet with coding vector $(1, 0, ..., 0)$ is delivered directly from the source to the receiver. Whether the relay has also delivered this particular uncoded packet successfully is irrelevant to studying the effect of using a systematic relay. Suppose the relay has successfully delivered the next three uncoded packets while the source failed to deliver them directly. If the relay were non-systematic, these three packets would have been coded, containing random coefficients similar to those in the fifth and sixth packets. In other words, using a systematic relay allows the receiver to obtain three additional uncoded packets than it would have otherwise. 

For a given set of erasure rates $P_1$, $P_2$, and $P_3$, we can compute the expected value of $U_{sys} - U_{non-sys}$ analytically:
\begin{align}
E[U_{sys} - U_{non-sys}] = M(P_1(1-P_2)(1-P_3)). \label{eq:EU}
\end{align}

\section{Simulations}\label{sec:simulations}
\begin{figure*}[t!]
\begin{center}
\subfloat[varying $P_1$ with $P_2 = P_3 = 0.2$.]{\label{fig:completion_p1}\includegraphics[width=.32\textwidth]{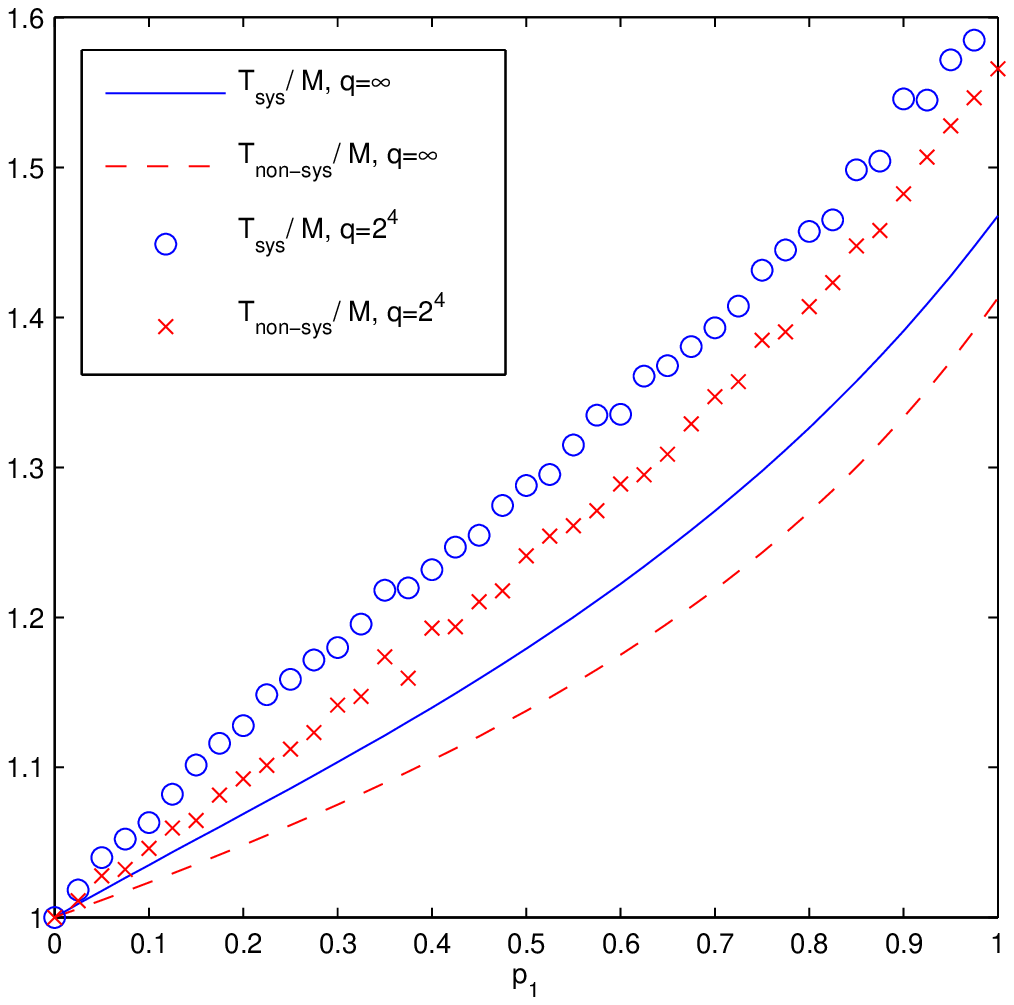}}\hspace*{.1cm}
\subfloat[varying $P_2$ with $P_1 = P_3 = 0.2$.]{\label{fig:completion_p2}\includegraphics[width=.32\textwidth]{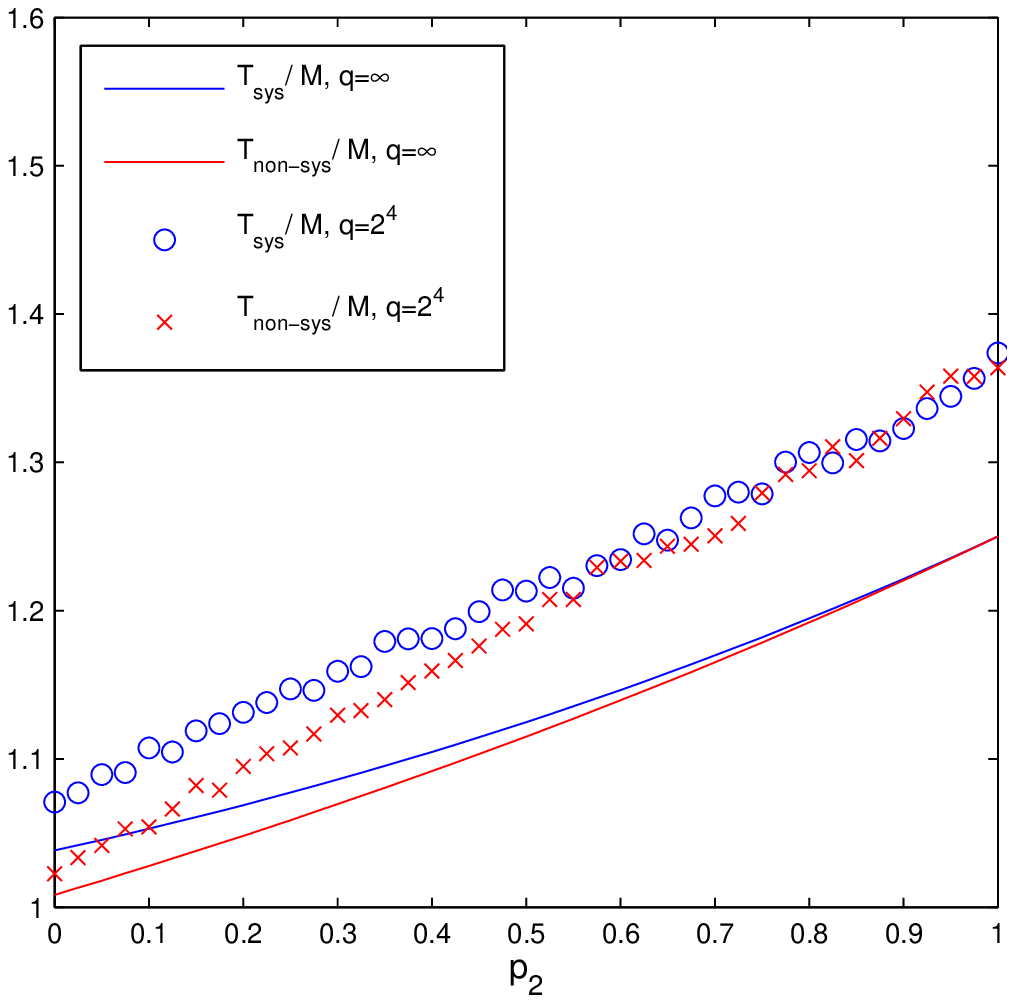}}\hspace*{.1cm}
\subfloat[varying $P_3$ with $P_1 = P_2 = 0.2$.]{\label{fig:completion_p3}\includegraphics[width=.32\textwidth]{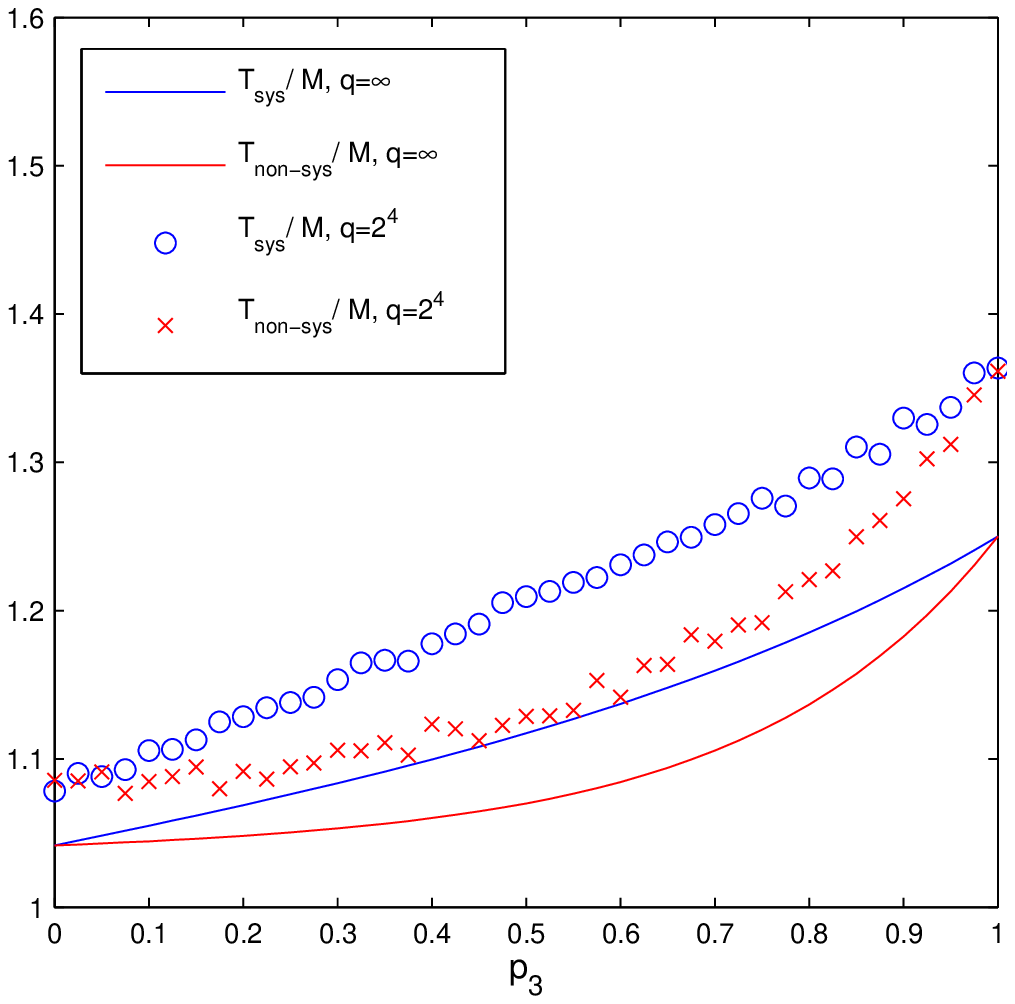}}
\end{center}
\caption{The mean completion time with systematic and non-systematic relays. In these figures, we present $T_{sys}/M$ and $T_{non-sys}/M$, i.e. mean completion time normalized by the number of packets $M$.}\label{fig:completiontime}\vspace*{-.1cm}
\end{figure*}

\begin{figure*}[t!]
\begin{center}
\subfloat[varying $P_1$ with $P_2 = P_3 = 0.2$.]{\label{fig:uncoded_p1}\includegraphics[width=.32\textwidth]{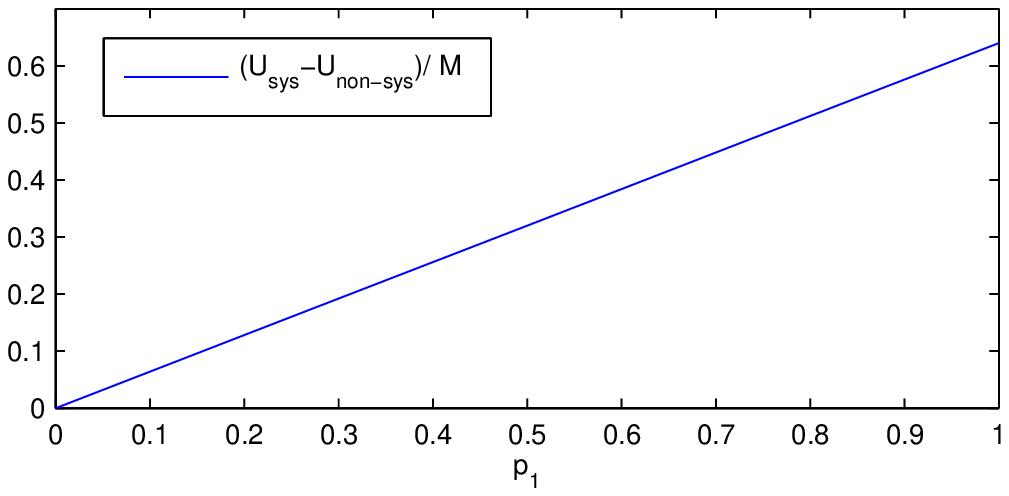}}\hspace*{.1cm}
\subfloat[varying $P_2$ with $P_1 = P_3 = 0.2$.]{\label{fig:uncoded_p2}\includegraphics[width=.32\textwidth]{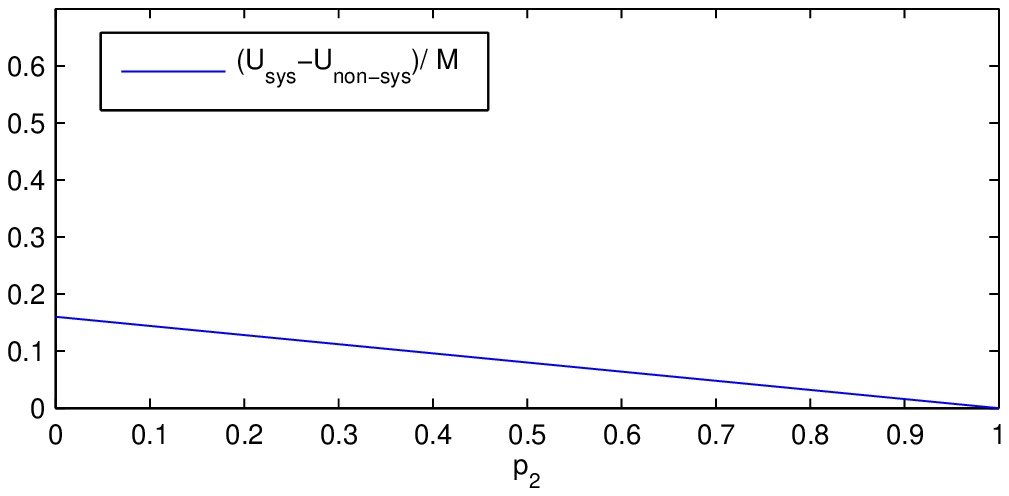}}\hspace*{.1cm}
\subfloat[varying $P_3$ with $P_1 = P_2 = 0.2$.]{\label{fig:uncoded_p3}\includegraphics[width=.32\textwidth]{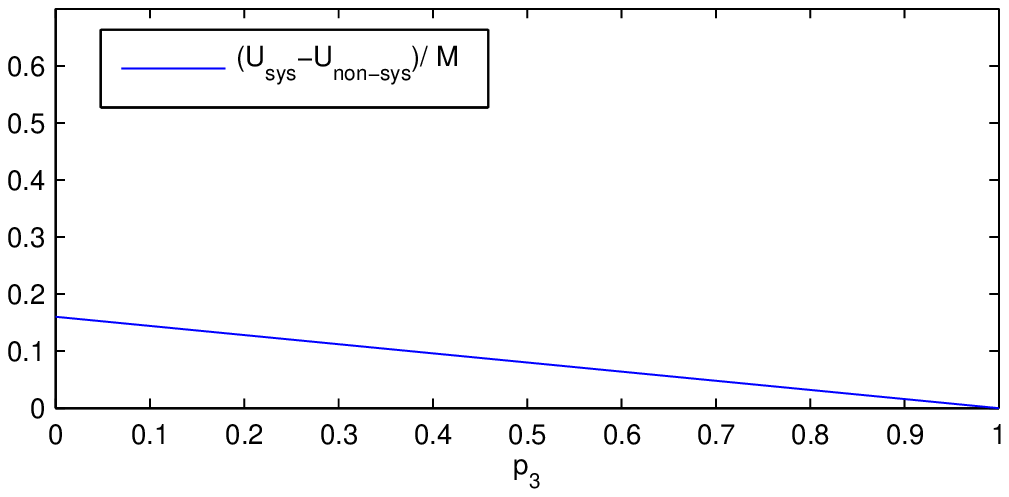}}
\end{center}
\caption{The effect of using a systematic relay. The figures illustrate the expected numbers of uncoded packets received via the systematic relay, $E[U_{sys} - U_{non-sys}]$, normalized by $M$. In essence, the figures illustrate the fraction of additional uncoded packets received at the receiver when using a systematic relay against when using a non-systematic relay.}\label{fig:uncodedpacket}\vspace*{-.3cm}
\end{figure*}

We use MATLAB to compute the mean completion times $T_{sys}$ and $T_{non-sys}$ given by Eq.~\eqref{eq:TNS} and \eqref{eq:TS}, as well as $E[U_{sys}-U_{non-sys}]$ which represents the gain in decoding complexity when using a systematic relay. As noted previously in Section~\ref{sec:model}, our analysis has assumed operations under an infinite field size $q=\infty$. Such an assumption ensures coded packets generated are linearly independent thus innovative and provide the best performance. We assume $M = 8$. Since our analysis does not capture explicitly the effect of field size on system performance, we provide simulation results for $q=2^4$. 
 
Figure~\ref{fig:completiontime} compares the mean completion times $T_{sys}$ and $T_{non-sys}$, when the value of a given $P_i$ is varied, $i\in \{1,2, 3\}$, while $P_j = 0.2$, $j \ne i$. This shows the effect of a given point-to-point link to the overall system performance. We normalize the mean completion times by $M$.  It can be observed that the performance differences between the non-systematic and the systematic relays follow similar trends regardless of the field size. In addition, there is a general increase in the expected completion time when the field size is small. This is because all coded packets are innovative when $q = \infty$, whereas coded packets over a finite field size have a non-zero probability to be linearly dependent. Nonetheless, the system with finite field size performs close to that of infinite field size.

Figure \ref{fig:uncodedpacket} compares the expected number of uncoded dofs at the receiver, $E[U_{sys}$] and $E[U_{non-sys}]$. We plot $\frac{E[U_{sys}-U_{non-sys}]}{M}$ to show the fraction of packets the receiver received uncoded from the systematic relay.

Figure~\ref{fig:completion_p1} shows the effect of the direct link between the source and the receiver for both $q=\infty$ and $q=2^4$. As the erasure rate $P_1$ increases, both $\frac{T_{sys}}{M}$ and $\frac{T_{non-sys}}{M}$ increase. Observe that, when $P_1 = 0$, the link between the source and the relay is reliable; thus, the relay provides no benefit regardless of whether it is systematic or not. However, as $P_1$ increases, the relay plays a bigger role in packet delivery to the receiver. The non-systematic relay optimizes for throughput instead of decoding complexity, hence performing better than the systematic relay, which occasionally transmits uncoded packets already received directly from the source. With $P_2$ and $P_3$ constant, the expected number of successfully transmitted, non-innovative, uncoded packets from the relay is approximately constant. This effect is illustrated by the relatively constant gap between the two curves $\frac{T_{sys}}{M}$ and $\frac{T_{non-sys}}{M}$. 

Corresponding to Figure~\ref{fig:completion_p1}, Figure~\ref{fig:uncoded_p1} plots $\frac{E[U_{sys}-U_{non-sys}]}{M}$ for the same parameters ($P_2 = P_3 = 0.2$). Observe that, when $P_1=0$, the relay provides no benefit regardless of whether it is systematic or not. As $P_1$ increases, $E[U_{sys}-U_{non-sys}]$ grows linearly according to Equation~\eqref{eq:EU}. When $P_1 = 1$, the source has to completely rely on the relay to deliver packets to the receiver. As a result, the receiver is able to receive uncoded packets when the relay is systematic; otherwise, the receiver only receives coded packets. With $P_2 = P_3 = 0.2$, the fraction of uncoded packets received at the receiver is $0.64$ when using a systematic relay.

From Figures~\ref{fig:completion_p1} and \ref{fig:uncoded_p1}, we observe that, although $\frac{T_{sys}}{M}$ is consistently larger than $\frac{T_{non-sys}}{M}$, the use of a systematic relay may present a reduction in decoding complexity. This reduction may be significant when the direct link between the source and the receiver is poor. This observation agrees with the intuition that relay plays an increasingly important role when the source has difficulty communicating with the receiver directly. Also note that, since decoding complexity is on the order of $O((M-U)^3)+O(U(M-U))$, the reduction in decoding complexity may be amplified as $M$ increases.

Figure~$\ref{fig:completion_p2}$ compares $\frac{T_{sys}}{M}$ and $\frac{T_{non-sys}}{M}$ when $P_1 = P_3 = 0.2$, while $P_2$ varies; Figure~\ref{fig:uncoded_p3} presents the corresponding $\frac{E[U_{sys}-U_{non-sys}]}{M}$. As $P_2$ increases, the relay's ability to aid the source degrades since it is less likely to receive successfully from the source. As a result, $\frac{T_{sys}}{M} = \frac{T_{non-sys}}{M}$ when $P_2 = 1$. On the other hand, when $P_2 = 0$, the relay receives everything the source transmits, thus it functions as a secondary source. Therefore, the systematic relay may repeat some uncoded packets while the non-systematic relay transmits innovative coded packets. In Figure~\ref{fig:uncoded_p3}, the value of $\frac{E[U_{sys}-U_{non-sys}]}{M}$ decreases with $P_2$, achieving a maximum value of $P_1(1-P_2)(1-P_3) = 0.16$ at $P_2=0$. In other words, when the channel between the source and the receiver is good, there is limited gain in using the relay systematically, because most of the uncoded packets are delivered directly from the source to the receiver. This observation is consistent with that from Figures~\ref{fig:completion_p1} and \ref{fig:uncoded_p1}.

In the last set of Figures~$\ref{fig:completion_p3}$ and \ref{fig:uncoded_p3},  $P_1 = P_2 = 0.2$, while $P_3$ varies. When $P_3=0$, any packet the relay receives is also available to the receiver. Therefore, whether the relay codes or not is irrelevant. As $P_3$ increases, the link between the relay and the receiver deteriorates; how effectively the relay utilizes opportunities to transmit innovative packets has a bigger effect on the completion time. As $P_3$ approaches 1, the relay is disconnected from the receiver; therefore, $\frac{T_{sys}}{M}$ and $\frac{T_{non-sys}}{M}$ converge. Similarly, in Figure \ref{fig:uncoded_p3}, we observe that as fewer innovative packets traverse through the relay to the receiver, the reduction in decoding complexity is diminished.

\section{Conclusion}\label{sec:conclusions}
We analyzed the problem of network coding in a relay channel, characterizing the delay performance and average decoding complexity at the receiver when the source relies on systematic network coding, and the relay tries to preserve this structure.
We focus on two specific policies for the relay: one that maintains the systematic structure and one that always generates random linear network coded packets with the information available in its buffer. 

We also proposed a metric to evaluate the decoding complexity advantage of relays that preserve the systematic structure. This metric accounts for the expected numbers of uncoded packets received via the systematic relay. 
Our numerical results show that, although maintaining a systematic structure comes at a small increase in delay in some scenarios, it can provide an important increase in the number of uncoded packets delivered to the destination, which translates in a lower decoding effort. 

\bibliography{networkcoding}
\bibliographystyle{IEEEtran}

\end{document}